\documentclass[prb,twocolumn,amssymb,floatfix]{revtex4-1}

\usepackage{graphicx}
\usepackage{dcolumn}
\usepackage{amsmath}
\usepackage[latin1]{inputenc}

\begin{document}
\title{Uncompensated magnetization and exchange-bias field in
La$_{0.7}$Sr$_{0.3}$MnO$_3$/YMnO$_3$ bilayers: The influence of the
 ferromagnetic layer}
\author{C. Zandalazini}
\altaffiliation[Permanent address: ]{FaMAF-CLCM at University of
Cordoba, Medina Allende S/N, 5000 Cordoba, Argentina.}
\affiliation{ Division of Superconductivity and Magnetism,
University of Leipzig, D-04103 Leipzig, Germany}
\author{P. Esquinazi}
\email[ ]{esquin@physik.uni-leipzig.de}\affiliation{ Division of
Superconductivity and Magnetism, University of Leipzig, D-04103
Leipzig, Germany}

\author{G. Bridoux}\altaffiliation[Present address: ]{Institut Catala de
Nanotecnologia (ICN), Universitat Autonoma de Barcelona, E-08193
Bellaterra, Spain} \affiliation{ Division of Superconductivity and
Magnetism, University of Leipzig, D-04103 Leipzig, Germany}

\author{J. Barzola-Quiquia} \affiliation{ Division of Superconductivity and Magnetism,
University of Leipzig, D-04103 Leipzig, Germany}
%%%%%%%%%%%%
%% Added by HO, feel free to rearrange
%%%%%%%%%%%%
\author{H. Ohldag}
\address{Stanford Synchrotron Radiation Lightsource, Stanford University, Menlo Park, CA 94025, USA}
\author{E. Arenholz}
\address{Advanced Light Source, Lawrence Berkeley National Laboratory, Berkeley, CA 94720, USA}
%%%%%%%%%%%%
%% End HO
%%%%%%%%%%%%

\date{\today}

\begin{abstract}
We studied the magnetic behavior of bilayers of multiferroic
 and nominally antiferromagnetic o-YMnO$_3$ (375~nm thick) and ferromagnetic
La$_{0.7}$Sr$_{0.3}$MnO$_3$ and La$_{0.67}$Ca$_{0.33}$MnO$_3$ ($8
\ldots 225~$nm), in particular the vertical magnetization shift
$M_E$ and exchange bias field $H_E$ for different thickness and
magnetic dilution of the ferromagnetic layer at different
temperatures and cooling fields. We have found very large $M_E$
shifts equivalent to up to 100\% of the saturation value of the
o-YMO layer alone. The overall behavior indicates that the
properties of the ferromagnetic layer contribute substantially to
the $M_E$ shift and that this does not correlate straightforwardly
with the measured exchange bias field $H_E$.
\end{abstract}


\maketitle

\section{Introduction}

In bilayers composed of antiferromagnetic (AFM) and a
ferromagnetic (FM) phases a ``horizontal" shift in the field axis
of the hysteresis loops is generally observed after cooling them
in a field applied at temperatures between the N\'eel $T_N$ and
Curie $T_C$ temperatures \cite{mei57,nog99}. This ``exchange-bias
field" $H_E$ has been studied in different systems due to its
fundamental importance as well as its technological relevance in
spin-valve sensors, actuators and in high-density recording media
\cite{sku03} and some details of the origin of $H_E$ are still a
matter of discussion.\cite{nog99}

Less studied is the shift in the magnetization axis, i.e. the
``vertical" $M_E$ shift in the hysteresis loop, probably because
of its rather small relative values \cite{mil00,hon06} and its
dependence on the cooling field $H_{\rm FC}$.\cite{nog00,kel02}
Recently, a maximum shift of 16\% of the saturation magnetization
was found in Fe$_x$Ni$_{1-x}$F$_2$/Co bilayers,  which appeared to
have an exchange bias field of  its own.\cite{che07} It was
proposed that $M_E$ is related to uncompensated moments (UCM) at
the AFM/FM interface and should have a direct correlation to
$H_E$.\cite{che07,now02prb} Element specific x-ray magnetic
studies of FeF$_2$/Co \cite{ohlpin,ohlfef} and CoO/Fe
\cite{gruy08} layered structures confirmed the existence of this
$M_E$ shift and revealed its relation to specific UCM in the AFM
material.

Due to the limited number of studies on the $M_E$ effect it is of
general interest to find systems with larger magnetization shifts,
not only because of its fundamental interest but also because this
shift provides a new degree of freedom in the hysteresis loop that
may be well have some applicability in future devices. In this
work we studied the exchange-bias shifts $H_E$ and $M_E$ of the
hysteresis loops as a function of temperature $T$ and $H_{\rm FC}$
for three AFM/FM bilayers having the same AFM layer but different
thickness and dilution of the FM layer. We observed an unusually
large uncompensated magnetization shift $M_E$ that is not simply
correlated with $H_E$ and does not originate only from the AFM
layer but from the FM one.

\section{Sample preparation details and x-ray characterization}

We prepared bilayers composed of a FM La$_{0.7}$Sr$_{0.3}$MnO$_3$
(LSMO) layer (selected for its weak anisotropy and small
coercivity) covering an AFM orthorhombic o-YMnO$_3$ (YMO) layer
grown on (100)~SrTiO$_3$ substrates of area $5 \times 5 ~$mm$^2$
for samples A and B and $6 \times 6 ~$mm$^2$ for sample~C. For the
depositions a KrF excimer laser (wavelength 248~nm, pulse duration
25~ns) was used and the optimal parameters found for o-YMO were
1.7~J/cm$^2$ with 5~Hz repetition rate, 800$^\circ$C and 0.10~mbar
for the substrate temperature and oxygen pressure during
preparation.  We have measured three bilayers, all of them with
the same 375~nm thick o-YMO layer on STO substrates prepared
always under the above mentioned conditions. To check the
reproducibility of the found effects we have prepared a fourth
bilayer with identical thickness as in sample A but instead of the
LSMO FM layer we used La$_{0.67}$Ca$_{0.33}$MnO$_3$ (LCMO)
deposited YMO and this last one on a (100)LSAT substrate.

For the FM LSMO layer, deposited immediately after the o-YMO one,
the parameters were 10~Hz repetition rate and 0.35~(0.38)~mbar
oxygen pressure, 8~(30)~nm thickness and at the same laser fluency
and substrate temperature, for sample A (B). In order to
corroborate the contribution of the FM layer in the $M_E$-shift we
have decreased further the oxygen concentration to  deposit the
LSMO film in sample~C (oxygen pressure 0.10~mbar) with a larger
thickness of 225~nm decreasing in this way its coercivity. For the
fourth LCMO/YMO  bilayer the YMO layer was grown under similar
conditions as before but the LCMO layer under an oxygen pressure
of 0.55~mbar; all other conditions as for the LSMO layers.

\begin{figure}[]
%\vspace{0.5cm}
\begin{center}
\includegraphics[width=1\columnwidth]{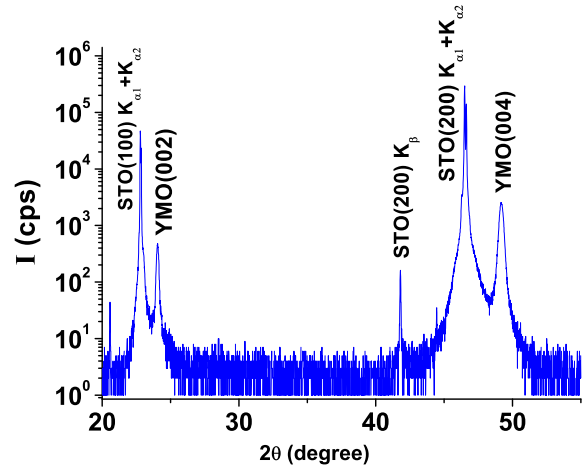}
\caption[]{X-ray spectrum of the single YMO AFM layer on STO
substrate.} \label{x-ray-ymo}
\end{center}
\end{figure}

\begin{figure}[]
%\vspace{0.5cm}
\begin{center}
\includegraphics[width=1\columnwidth]{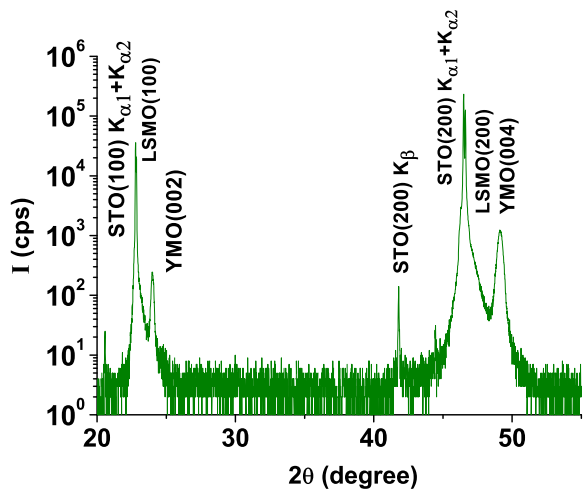}
\caption[]{X-ray spectrum of the bilayer sample~B. The labels
indicate the corresponding the main diffraction peaks.}
\label{x-rb}
\end{center}
\end{figure}

The epitaxial growth in the ${00l}$ direction for the o-YMO and
${l00}$ for LSMO phases was confirmed by x-ray diffraction using
Cu-K$_\alpha$ line. As an example we show in Fig.~\ref{x-ray-ymo}
the the x-ray spectrum of the single o-YMO layer on STO. The
preferential growth of the $(00l)$ planes of the orthorhombic
phase YMO is clearly seen. Within the experimental resolution no
maxima due to the hexagonal phase are observed. Figure \ref{x-rb}
shows the x-ray spectrum obtained for sample~B. The main
diffraction peaks from the LSMO layer are observed as a weak
shoulder near the STO main maxima. Magnetization measurements were
performed with a superconducting quantum interference device
(SQUID) from Quantum Design in the temperature range between 5~K
and 350~K.

%%%%%%%%%%%%%%%%%%%%%%%%%%%%%%%%
%% Added by HO
%%%%%%%%%%%%%%%%%%%%%%%%%%%%%%%%

In addition,  we performed soft x-ray absorption and circular
dichroism measurements using the bending magnet beamline 6.3.1 at
the Advanced Light Source in Berkeley, CA (USA) and the elliptical
undulator beamline 13.1 at the Stanford Synchrotron Radiation
Lightsource, Stanford, CA (USA). For these measurements the sample
was mounted between the poles of an electromagnet so that the
x-rays are incident on the sample under a grazing angle of
30$^\circ$ parallel to the direction of the applied magnetic
field. The x-ray absorption intensity was monitored using the
electron yield method. Hysteresis loops were acquired by sweeping
the external field while monitoring the electron yield at the Mn
L$_3$ and L$_2$ absorption resonance ($\approx$ 640~eV). This
approach is surface sensitive and in general it yields information
only on the first $ \sim 5~$nm of the sample. Assuming an
exponential escape depth of 2.5~nm, then 95\% of the signal comes
from the top 6~nm of the sample. This is essentially our probing
depth. For a more detailed description of the technique see
Refs.~\onlinecite{ohlpin,ohlfef}.

%%%%%%%%%%%%%%%%%%%%%%%%%%%%%%%%
%% End HO
%%%%%%%%%%%%%%%%%%%%%%%%%%%%%%%%

\section{Results}
\subsection{Single YMnO$_3$ layers}

According to literature\cite{hsi08,kim06} the o-YMO phase is AFM
with N\'eel temperature $T_N = 42 \pm 2~$K  and with a
ferroelectric transition at $\sim 31~$K. In spite of its low $T_N$
this material has several advantages for exchange bias studies. It
belongs to the family of the perovskite manganite RMnO$_3$  and
the magnetic and electrical properties can be changed by cation
substitution keeping similar lattice constants and therefore
without drastic changes in its structural properties. On the other
hand, o-YMO is a phase that was not thoroughly studied yet and the
influence of its ferroelectric behavior, in spite of the low
temperature, might be used as a paradigm for potential
applications in magnetoelectric devices.\cite{lau06}

\begin{figure}[]
%\vspace{0.5cm}
\begin{center}
\includegraphics[width=1\columnwidth]{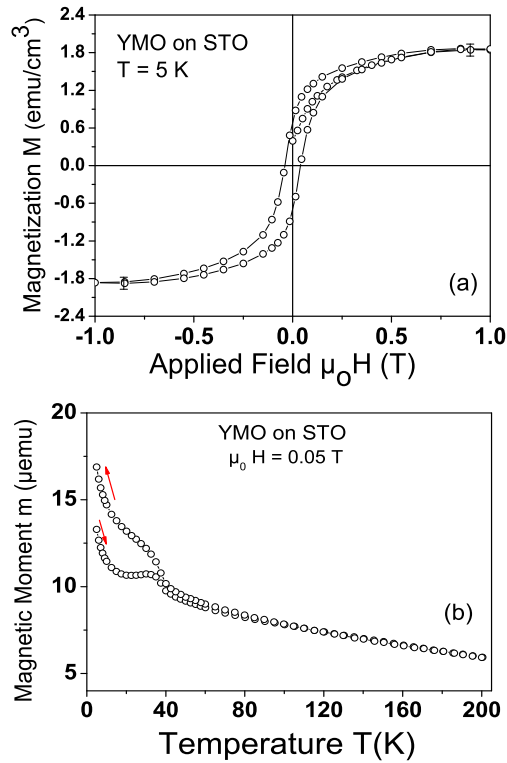}
\caption[]{(a) Hysteresis loop of the magnetization at 5K for the
375~nm thick YMO layer on STO. The error bars indicate the maximum
error due to the SQUID and geometry measurements. (b) Temperature
dependence of the magnetic moment of a single YMO layer on STO in
ZFC and FC states at an applied field of 0.05~T. An error bar of
$\pm 0.3~\mu$emu is the expected maximum error from our SQUID
measurements.} \label{mymo}
\end{center}
\end{figure}

Figure~\ref{mymo}(a) shows the magnetization loop of single o-YMO
layer. The hysteresis loop indicates a magnetization at saturation
of 1.8~emu/cm$^3$ at 5~K and at applied fields $\mu_0 H > 0.5$~T
in agreement with reported values.\cite{li09} Figure~\ref{mymo}(b)
shows the magnetic moment of a single o-YMO layer ($6 \times 6
\times 0.375~10^{-3}$~mm$^3$) on STO measured as a function of
temperature in ZFC and FC states at an applied field of 0.05~T. A
clear increase in $m(T)$ decreasing temperature is observed at $T
\simeq 42~$K. An hysteresis between ZFC and FC is observed already
below $T \sim 60~$K. As was shown in earlier studies on YMO  we
may expect to have persistent spin waves at temperatures above
$T_N$.\cite{dem07}

From the hysteresis loop shown in Fig.~\ref{mymo}(a) one may
speculate that the YMO film behaves as a ferro- or ferrimagnet and
not as an antiferromagnet. In fact, a recent study  suggests a
change of the usual bulk antiferromagnetic state to a
strain-dependent non-collinear magnetic one in thinner ($\lesssim
120~$nm) o-YMO films.\cite{mar09} Taking into account that our YMO
layers are much thicker and show a different $m(T)$ behavior (at
ZFC and low applied fields the measured $m(T)$ of our YMO films
alone resembles practically the usual $T-$dependence found for
antiferromagnets) as those reported in Ref.~\onlinecite{mar09} we
remark that the magnetic behavior of the o-YMO layers may
correspond to the one observed in diluted antiferromagnets in
external magnetic field (DAFF). It is well known that DAFF develop
a domain state when cooled below $T_N$ (sometimes with a
spin-glass-like behavior) and this leads to a net magnetization,
which couples to the external field, see e.g.
Refs.~\onlinecite{mil00,kel02,now02,rad03,zab08}.

From the measured temperature dependence of the magnetic moment
and the observed scaling of the exchange bias field $H_E$ with the
inverse of the thickness of the LSMO layer for samples A and B,
see section~\ref{bi1}, and the quantitative agreement of the
obtained $H_E$ and $M_E$ shifts for the fourth  sample (similar to
sample A but with LCMO instead of LSMO) we may conclude that YMO
behaves as an AFM or DAFF layer for the exchange bias effects.
Whatever the real magnetic equilibrium state of our o-YMO films
is, we may expect to see exchange bias effects when these films
are coupled to a ferromagnet. Further examples for exchange bias
effects in heterostructures with different ferro- or ferrimagnets
can be seen in Refs.~\onlinecite{fit09,ke04} and $H_E$ effects,
positive as well as negative, has been also observed in
ferrimagnetic based bilayers.\cite{can01}

\subsection{La$_{0.7}$Sr$_{0.3}$MnO$_3$/YMnO$_3$ bilayers}
\label{bi1}

\begin{figure}[]
%\vspace{0.5cm}
\begin{center}
\includegraphics[width=1\columnwidth]{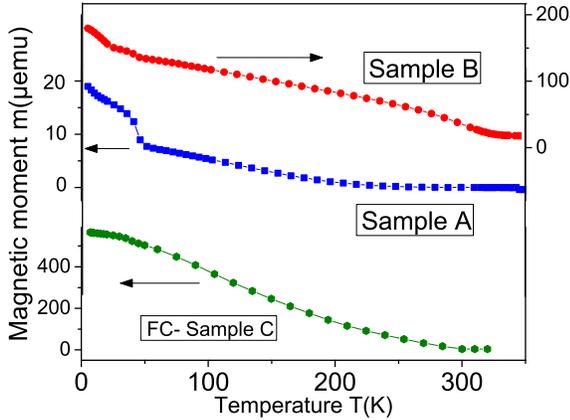}
\caption[]{Temperature dependence of the remanence for samples A and
B measured at zero field after cooling them to 5~K at 0.1~T in-plane
field. Also shown is the field cooling curve at 0.1~T for sample C.
Note the difference scales of the $y-$axis for each sample.}
\label{mrvst}
\end{center}
\end{figure}

Figure~\ref{mrvst} shows the remanent moment for samples A and B
measured increasing temperature at zero field, after cooling them
to 5~K in a field of 0.1~T applied in-plane, i.e. $a$ or $b$
direction. Changes in slope of the remanence moment are observed
near the N\'eel temperature onset $T_N \sim 50~$K of the o-YMO
layer. This increase of $\sim 8~$K in $T_N$ might be related to
the an exchange-bias\cite{lie07,zaa00} or strain\cite{shi04}
effect. An anomaly is also observed at $T \sim 20~$K, as shown in
Fig.~\ref{mymo}(b), and already reported in the
literature.\cite{hsi08,fin10} The temperature dependence of the
remanence measured in sample~B shows a clear change of slope near
the Curie temperature of the LSMO layer. In contrast, due to the
smaller LSMO thickness the remanent moment of sample~A does not
show a clear anomaly at $T_C$; similarly for sample~C (not shown).
For sample~C we show in Fig.~\ref{mrvst} the field cooled (FC)
curve at 0.1~T; the absence of a marked anomaly at $T_C$ and the
smooth decrease of the magnetic moment with $T$ demonstrates the
expected strong magnetic dilution of the LSMO film. The existence
of the FM state in this layer was confirmed through hysteresis
loop measurements up to its ferromagnetic onset at $T_C \sim
300~$K. The FC results presented below were obtained always after
cooling the samples from $T > T_C$ at zero field and after
applying an in-plane field $H_{\rm FC}$ at 100~K~$> T_N$.

% effect of the interface in the hysteresis

\begin{figure}[]
%\vspace{0.5cm}
\begin{center}
\includegraphics[width=1\columnwidth]{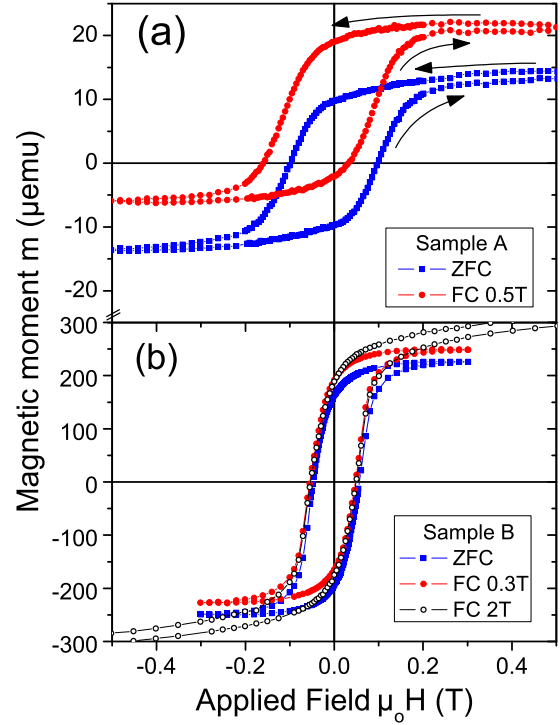}
\caption[]{Hysteresis loops at 5~K measured for samples~A (a) and B
(b) after zero-field cooled (ZFC) and field cooled (FC) states  at
the fields shown in the figures. The arrows indicate the sweeping
field direction starting the loop always from positive fields.}
\label{2}
\end{center}
\end{figure}

Figures~\ref{2}(a) and (b) show the hysteresis loops for ZFC and
FC measurements at 5~K for samples A and B.  A remarkable $M_E$
shift of the same order of the saturation magnetic moment $m_s$ is
observed for sample~A after FC from 100~K at $\mu_0 H_{\rm
FC}=0.5$~T. For sample~B the $M_E$ shift is also clearer measured
but it is smaller relative to $m_s$. The sign of the $M_E$-shift
changes when the direction of $H_{\rm FC}$ changes, i.e. it has
the same sign as that of $H_{\rm FC}$. This indicates that the
effective UCM layer is pinned in the direction of the cooling
field, which means a ferromagnetic coupling.

In the determination of the $M_E$ and $H_E$ shifts we took special
care to rule out effects due to minor hysteresis loops\cite{pi06}.
Studying the behavior of the loops at different $H_{\rm FC}$ we
conclude that no minor loops and a clear saturation behavior of
the magnetic moment are obtained for $\mu_0 H_{\rm FC} \geq 0.2~$T
at $T \geq 5~$K for samples A and B, see Fig.~\ref{2}. For
sample~C, which has a more diluted and inhomogeneous FM layer, the
hysteresis  loops reveal no complete saturation at $\mu_0 H_{\rm
FC} < 0.4~$T. However minor loop effects can be neglected also for
this sample at $\mu_0 H_{\rm FC} \geq 0.2~$T, as the behavior of
the coercive field $H_c$ vs. $H_{\rm FC}$ indicates (see
Fig.~\ref{4} below).

We note that the value of $m_s$ obtained from the hysteresis loops
depends on the applied $H_{\rm FC}$. As example we show this
effect for sample~B where the hysteresis loop was measured after
cooling the sample at $\mu_0 H_{\rm FC} = 2$~T, see
Fig.~\ref{2}(b). This effect is due to the LSMO layer and
indicates that the number of aligned domains can be changed with
$H_{\rm FC}$. In this case we expect that the $M_E$ effect will be
strongly influenced by the FM layer since, as in the case of a
diluted AFM layer\cite{now02prb,kel02}, the formation and number
of its domains that take part in the exchange bias coupling with
the AFM layer can be enhanced leading to an increase of $M_E$.
Note however that the $M_E$ effect is expected to decrease with
$H_{\rm FC}$, i.e. $M_E \rightarrow 0$ for $H_{\rm FC} \rightarrow
\infty$.

Note the opening of $\sim 1~\mu$emu of the hysteresis at the end of
the loop at 0.5~T for sample~A, see Fig.~\ref{2}(a). A similar
opening is measured for all samples in agreement with the numerical
results obtained with the domain state model for exchange bias
proposed by Nowak et al.\cite{now02prb,now02}.  The fact that the
loops do not close indicates that uncompensated spins - pinned
earlier during the field cooling - rotate and remain pinned in the
opposite direction during the field sweep loop, reducing the final
saturation moment. We note that in all three bilayers this opening
remains of the order of $1 \ldots 2~\mu$emu, i.e. several times
smaller than the $M_E$ shift, as we show below.

\begin{figure}[]
%\vspace{0.5cm}
\begin{center}
\includegraphics[width=1\columnwidth]{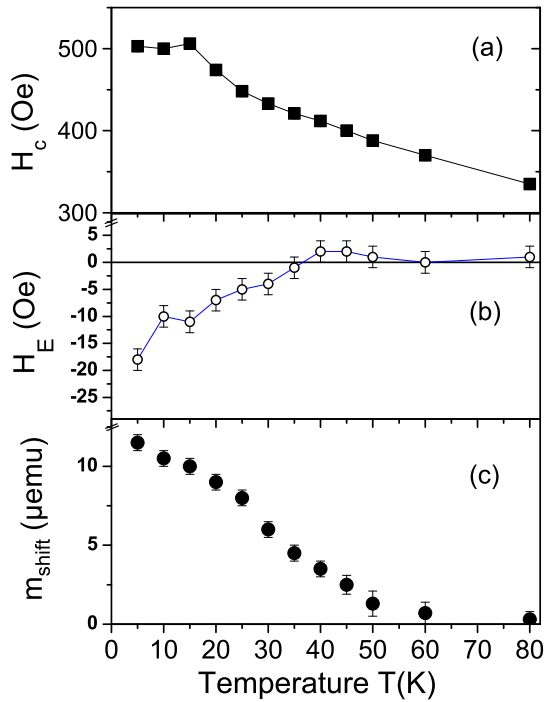}
\caption[]{Temperature dependence of the coercivity (a) and exchange
bias (b) fields and of the shift in magnetic moment $m_{\rm shift}$
due to the $M_E$ effect (c) for sample~B after cooling it in a field
of 0.3~T.} \label{3}
\end{center}
\end{figure}

\begin{figure}[]
%\vspace{0.5cm}
\begin{center}
\includegraphics[width=1\columnwidth]{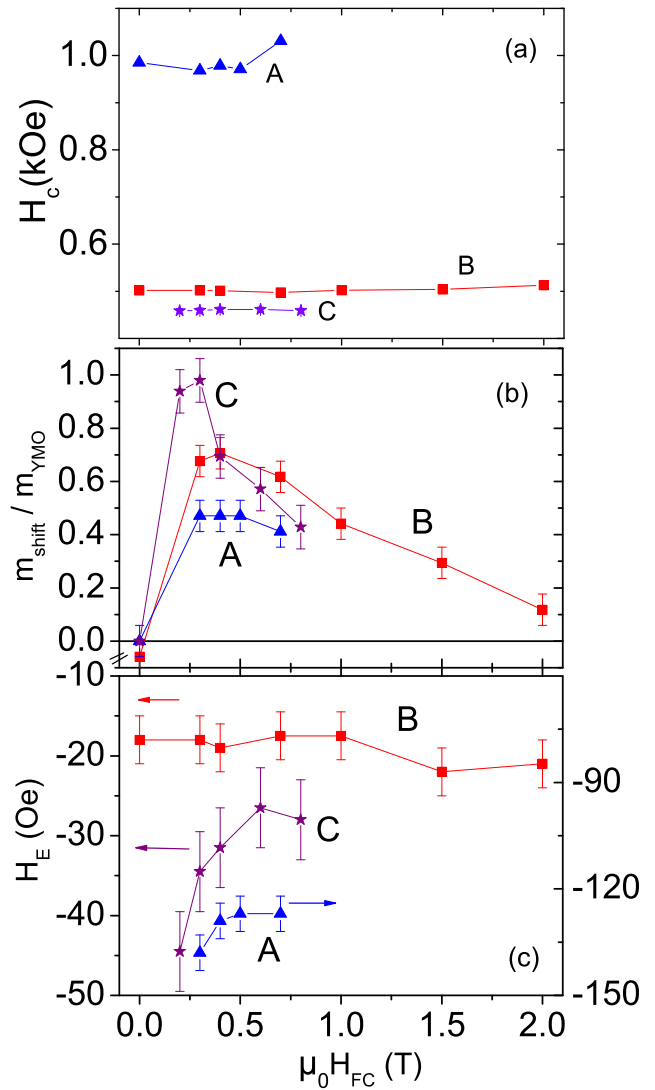}
\caption[]{Dependence of the coercive field (a), shift in magnetic
moment $m_{\rm shift}$ (b) and exchange bias field $H_E$ (c) on the
cooling field $H_{\rm FC}$ for the three measured bilayers at 5~K. In
(b) we plot $m_{\rm shift}$ normalized by the maximum saturation
moment $m_{\rm YMO}$ of the o-YMO layer, i.e. $m_{\rm YMO} =
17~\mu$emu for samples A and B and $24.5~\mu$emu for sample~C. Note
that the values of $m_{\rm shift} \sim 0$ at $H_{\rm FC} = 0$ were
obtained using maximum fields between 0.3 and 0.5~T for the
hysteresis loops. For all the other points the maximum field of the
loops coincides with $H_{\rm FC}$.} \label{4}
\end{center}
\end{figure}

To characterize quantitatively the exchange bias $M_E$ effect and
for a direct comparison with the saturation magnetic moments of
each of the layers we define it as $m_{\rm shift}=(m_s^+ +
m_s^-)/2$, where $m_s^+$ and $m_s^-$ are the saturation moments at
positive and negative fields. The shift in the field axis is
defined as $H_E=(H_c^+ + H_c^-)/2$, where $H_c^+$ and $H_c^-$ are
the coercive fields in upward and descending loop branches,
respectively. We note that the $H_E$ values were obtained only
after centering the hysteresis loop, subtracting the upward $M_E$
shift.

Figure~\ref{3} shows the coercivity $H_c$ (a), the exchange-bias
$H_E$ (b) and the vertical shift in magnetic moment $m_{\rm
shift}$ (c) as a function of $T \leqslant 80$~K for sample~B,
measured after $\mu_0 H_{\rm FC} = 0.3~$T, as an example. A
similar behavior is observed  for samples A and C. Both, $H_C$ and
$H_E$ show an anomaly at $T \lesssim 20~$K, in agreement with the
behavior found in the remanence curve, see Fig.~\ref{mrvst},
suggesting that the transition at that temperature influences the
exchange interaction. At $T \gtrsim 35~$K $H_E$ crosses zero and
changes to positive. This sign change of $H_E$ from negative to
positive increasing temperature was observed also in CoO/Co
bilayers\cite{rad03} and suggests  a change from direct ($J_{\rm
interface}>0$) to indirect ($J_{\rm interface} < 0$) interface
interaction. As expected, $H_E(T)$ as well as $m_{\rm shift}$
vanish at $T \gtrsim T_N$. In contrast to $H_E(T)$ no anomalous
behavior is observed in $m_{\rm shift}(T)$ at $T < T_N$, with
exception of the slope change at $T \sim 20$~K, see
Fig.~\ref{3}(c).

Figure~\ref{4} shows the $H_{\rm FC}$-dependence of $H_c$, $m_{\rm
shift}$ and $H_E$ for the three samples measured at 5~K.  The
decrease of $H_E$ from samples A to B agrees with the expected
inverse proportionality of $H_E$ with the thickness of the FM layer.
According to this thickness dependence  sample~C should show nearly
one order of magnitude smaller $H_E$ than for sample B, in clear
disagreement with the obtained result, see Fig.~\ref{4}(c),
suggesting that the magnetic dilution of this sample is responsible
for the large observed $H_E$ field.

Regarding the $M_E$ effect and in agreement with the results in
Co/CoO bilayers\cite{kel02} we observe a vanishing effect at zero and
at large enough values of $H_{\rm FC}$, see Fig.~\ref{4}(b). Under
the assumptions done in Refs.~\onlinecite{now02,now02prb} the $M_E$
shift is mainly due to the AFM layer. According to this model, the
largest $m_{\rm shift}$ expected from our o-YMO layer, assuming
complete saturation in the whole 375~nm thick layer, would be $m_{\rm
YMO} = 17~\mu$emu and $24.5~\mu$emu from samples A or B and C,
respectively. To estimate those numbers we have taken into account
the measured magnetization at saturation of the single layers. The
normalized $m_{\rm shift}$ by the corresponding $m_{\rm YMO}$, see
Fig.~\ref{4}(b), would indicate that it is necessary that 50\% to
70\% of the YMO layer should be responsible for the measured $m_{\rm
shift}$ at $H_{\rm FC} \sim 0.5~$T. This percentage increases further
for the diluter sample~C at $0.2~$T$ \leq H_{\rm FC} \leq 0.4~$T.
Taking into account the 375~nm thickness of the YMO layer this
assumption appears unlikely.

We remark that unexpected phenomena can occur at oxide interfaces.
A recent study, for example, found an excess magnetization
produced at the interface between STO and an AFM
La$_{1/3}$Ca$_{2/3}$MnO$_3$ layer,\cite{hof09} which origin
remains unclear. In our case the large  $m_{\rm shift}$ values --
actually a giant $M_E$ effect -- indicate that a large
contribution should come from the FM layer. Taking into account
the saturation moments of the LSMO layers alone, we estimate for
example that a thickness of the LSMO layer of less than 1.3~nm for
sample~B and $< 10$~nm for sample~C should be enough to produce
the observed $m_{\rm shift}$ at $H_ {\rm FC} = 0.5~$T.

\subsection{La$_{0.67}$Ca$_{0.33}$MnO$_3$/YMnO$_3$ bilayer}

\begin{figure}[]
%\vspace{0.5cm}
\begin{center}
\includegraphics[width=1\columnwidth]{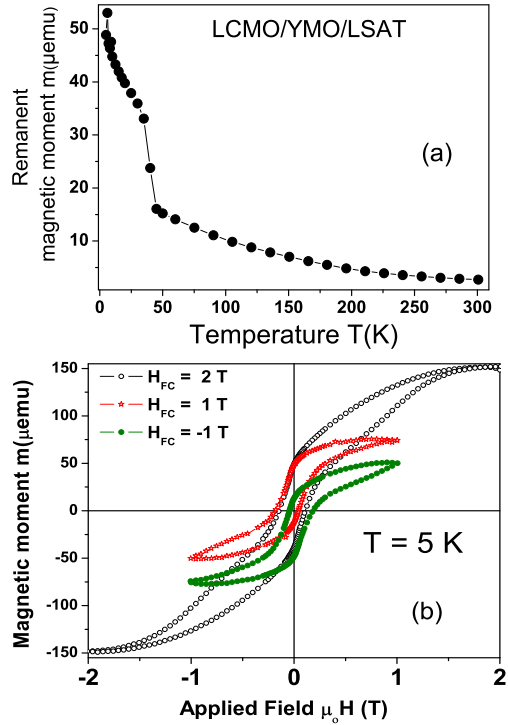}
\caption[]{(a) Temperature dependence of the zero field remanent
magnetic moment measured after field cooled at 1~T of a bilayer
La$_{0.67}$Ca$_{0.33}$MnO$_3$ (8~nm) / YMnO$_3$ (375~nm), similar
to sample A, but the YMO layer first deposited on a (100)LSAT
substrate. (b) Hysteresis loops at 5~K measured for the same
sample after field cooled (FC) at the fields shown in the figure.}
\label{lcmo-1}
\end{center}
\end{figure}

\begin{figure}[]
%\vspace{0.5cm}
\begin{center}
\includegraphics[width=1\columnwidth]{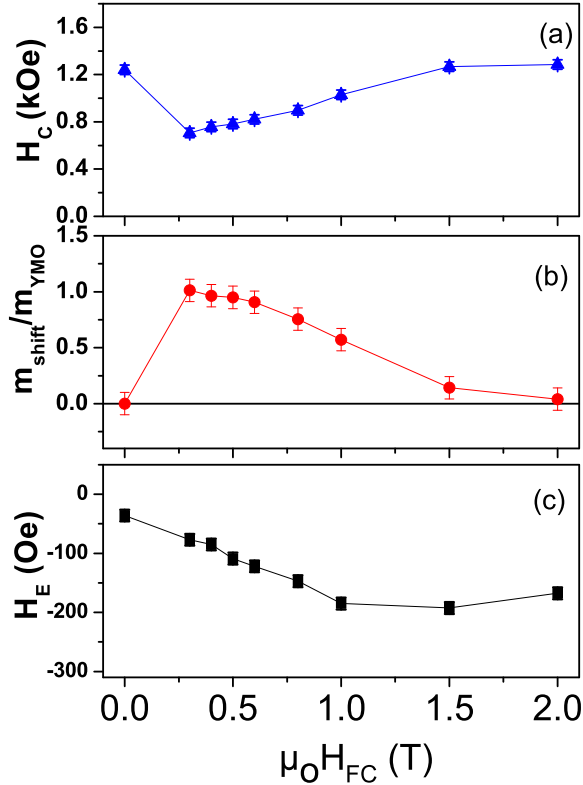}
\caption[]{Similar to Fig.~\protect\ref{4} but for the YMO/LCMO
bilayer: Dependence of the coercive field (a), shift in magnetic
moment $m_{\rm shift}$ normalized by the maximum saturation moment
$m_{\rm YMO}$ of the o-YMO layer alone (b), and exchange bias
field $H_E$ (c), on the cooling field $H_{\rm FC}$ at 5~K.}
\label{lcmo-2}
\end{center}
\end{figure}

Further evidence for the reproducibility and robustness of the
effects observed in the three LSMO/YMO bilayers reported in the
last section are provided by the results of a LCMO/YMO bilayer
with similar geometry and preparation conditions as sample A.
Figure~\ref{lcmo-1}(a) shows the remanent magnetic moment of this
bilayer after cooling the sample at 1~T applied field. The
transition at the N{\'e}el temperature of the YMO layer is clearly
seen as well as the change of slope at $\sim 20~$K. In
Fig.~\ref{lcmo-1}(b) the hysteresis loops for three field cooled
states at  fields $H_{\rm FC} = \pm 1~$T and 2~T are shown. At low
$H_{FC}$ fields the exchange bias $M_E$- and $H_E$-effects are
clearly observed whereas at high enough fields the $M_E$ effect
vanishes, see Fig.~\ref{lcmo-1}(b). Figure~\ref{lcmo-2} shows the
$H_{\rm FC}$-dependence for the three characteristics parameters.
The observed $m_{\rm shift}$ at $H_{\rm FC} \lesssim 1~$T, see
Fig.~\ref{lcmo-2}(b), is as large as the magnetic moment at
saturation of the 375~nm thick YMO layer alone,  indicating
clearly that the FM layer should contribute to this effect near
the interface.

Although in the LSMO/YMO bilayers we did not find any
correspondence between the coercive field $H_c(H_{\rm FC})$ and
$m_{\rm shift}(H_{\rm FC})$, see Fig.~\ref{4}, one may expect some
correlation between them in case of a bilayer with a very thin
(and diluted) FM layer. This may be so if we take into account the
amount of the FM layer that remains pinned at the interface. In
this case the smaller the effective thickness of the remained
unpinned ferromagnetic layer the smaller might be $H_c$.
Apparently this is observed in the (thin)LCMO/(thick)YMO bilayer.
Indeed, the results shown in Fig.\ref{lcmo-2} indicate that when
$m_{\rm shift}$ decreases at $H_{\rm FC} > 0.25$T, i.e. when the
amount of UCM decreases, $H_c$ increases.

\section{Discussion and Conclusion}

%%%%%%%%%%%%%%%%%%%%%%%%%
%% Added by HO
%%%%%%%%%%%%%%%%%%%%%%%%

\begin{figure}[]
%\vspace{0.5cm}
\begin{center}
\includegraphics[width=1\columnwidth]{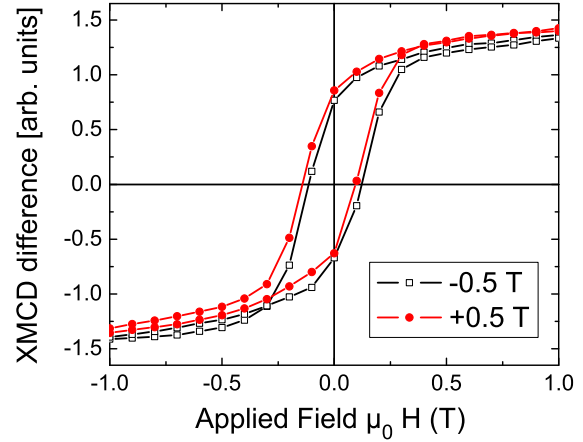}
\caption[]{Hysteresis loops of sample A acquired at 15~K after
cooling in a field of either +0.5~T or -0.5~T using x-ray magnetic
circular dichroism and the Mn L-absorption resonance. The loops
exhibit a horizontal loop shift H$_{\rm E} $ of 140~Oe as well as
a vertical shift $m_{\rm shift} \simeq 5\%$ of the saturation
value.} \label{xmcd}
\end{center}
\end{figure}

To further corroborate our conclusion that the observed  vertical
shift is mainly due to the FM and its interface region with the
AFM layer we show the hysteresis loops acquired using x-ray
magnetic circular dichroism in Fig.~\ref{xmcd}. For sample~A we
find a shift of about 5\% using the surface sensitive approach
measuring the response of the Mn ions within the LSMO FM layer
only. The observed vertical shift is a clear indication that the
FM layer is contributing to the $M_E$ effect and that the shift is
not confined to the bulk of the AFM. Assuming that 95\% of the
secondary electrons detected in our experiment originate from the
top 6~nm \cite{ohlpin,ohlfef} we can conclude that the interfacial
region of the FM/AFM layer contributes significantly more to the
m$_{\rm shift}$ compared to the surface layers of the FM. This
result agrees with the estimate from the bulk SQUID measurements
that one needs about 1~nm thick FM layer (for samples A as well as
B)  to account for the observed $m_{\rm shift}$. Taking into
account the previous statement that it is highly unlikely that the
entire AFM bulk contributes to the shift we can conclude that the
excess magnetization is produced predominately at the FM interface
during the field cooling process due to interfacial exchange
coupling between the AFM and the FM as shown previously for the
case of Co/FeF$_2$ \cite{ohlfef}.

%%%%%%%%%%%%%%%%%%%%%%%%%
%% Figure added by HO
%%%%%%%%%%%%%%%%%%%%%%%%%

%%%%%%%%%%%%%%%%%%%%%%%%
%% end HO
%%%%%%%%%%%%%%%%%%%%%%%%

Using similar arguments on the importance of the magnetic dilution
of the AFM layer \cite{now02,now02prb}, we argue that in our
system the dilution of the FM layer may play a mayor role in the
$M_E$ shift. In other words, the robust AFM layer influences the
magnetic behavior of the FM one, within a certain thickness from
the interface. We note that some kind of $M_E$-shift were recently
reported for ferrimagnetic very thin hard/soft (3nm/12nm)
DyFe$_2$/YFe$_2$ heterostructures.\cite{fit09} However, in that
work the $M_E$ effect is in opposite direction to that of the
applied $H_{\rm FC}$, in contrast to our observations.

Furthermore, a comparison between the overall behavior obtained
for $m_{\rm shift}(H_{\rm FC})$ and $H_E(H_{\rm FC})$ indicates
that there is no simple correlation between the two exchange bias
effects. Note that $H_E$ decreases strongly from sample A to B,
whereas $m_{\rm shift}$ increases. Although element selective
x-ray magnetic measurements would help to determine the
penetration depth of the UCM in each of the layers, it is clear
from our SQUID measurements that the o-YMO layer alone cannot be
the reason for the observed giant $M_E$ effect, this is the main
message of our work.

In conclusion, our studies on LSMO/o-YMO bilayers and on a single
LCMO/o-YMO bilayer found large uncompensated $M_E$ shifts, whose
sign correlates with the direction of the cooling field $H_{\rm
FC}$. Both, the exchange-bias $H_E$ and $M_E$ effects, vanish near
$T_N$ of the YMO layer. The large $m_{\rm shift}$ values indicate
that the AFM layer cannot be the only responsible but a certain
thickness of the FM layer near the interface.  This behavior can
be actually understood taking similar arguments as those used for
the AFM layer in the domain state exchange-bias model of
Refs.~\onlinecite{now02prb,now02}. Tuning the thickness and
magnetic dilution of the FM layer one should be able to obtain
large $M_E$ shifts making it an effect worth to study in systems
with $T_N
> 300~$K. The different behaviors of $H_E$ and $M_E$ with
temperature, cooling field and FM layer thickness indicate that these
two phenomena are not correlated in a simple way.

\acknowledgments We thank M. Ziese for fruitful discussions and
for pointing us Ref.~\onlinecite{pi06} and A. Setzer for technical
assistance. This work was supported by the DFG within the
Collaborative Research Center (SFB 762) ``Functionality of Oxide
Interfaces''. One of us (C.Z.) was supported by the S\"achsisches
Staatsministerium f\"ur Wissenschaft und Kunst under
4-7531.50-04-0361-09/1.
%%%%%%%%%%%%%%%%%%
%% added by HO
%%%%%%%%%%%%%%%%%%
SSRL and ALS are national user facilities supported by the
Department of Energy, Office of Basic Energy Sciences. SSRL is
operated by Stanford University and ALS is operated by the
University of California.
%%%%%%%%%%%%%%%%%
%% End HO
%%%%%%%%%%%%%%%%%

%\bibliography{D:/data/Zandalazini/bilayers}
%\bibliography{D:/DATA/Future_SFB/ZnO/light-induced/bilayers}

\end{document}